# The 2011 February superoutburst of the dwarf nova SDSS J112003.40+663632.4

Jeremy Shears, Steve Brady, Tom Krajci, Enrique de Miguel, Mike Potter, Richard Sabo and William Stein

## Abstract

We report unfiltered photometry of SDSS J112003.40+663632.4 during the 2011 February outburst which revealed the presence of superhumps with peak-to-peak amplitude of up to 0.22 magnitudes showing this to be an SU UMa type dwarf nova. The outburst amplitude was 5.4 magnitudes above mean quiescence and it lasted at least 12 days. The mean superhump period during the plateau phase was $P_{sh}$ = 0.07057(19) d.

## History and outbursts of SDSS J112003.40+663632.4

SDSS J112003.40+663632.4 was first identified as a dwarf nova by Wils *et al.* (1) during their search for cataclysmic variables in which they cross matched blue objects from the Sloan Digital Sky Survey (SDSS) with data from the Galaxy Evolution Explorer (GALEX) UV telescope. They reported that SDSS imaging revealed the object at r ≈ 21 and g = 21.1 in quiescence. Two outbursts were also detected by SDSS: in one outburst the object reached r = 15.9 (2000 May) and on a second occasion (2008 Jan) it was observed in decline from outburst at r = 18.2. At this time the SDSS spectrum showed a steep blue continuum with broad Balmer absorption lines, on which weak emission lines are superimposed with only the central core visible, consistent with an accretion disc in outburst. During quiescence the spectrum exhibits narrow, moderately strong Balmer emission lines superimposed on a weak continuum.

The Catalina Real-Time Transient Survey (CRTS) detected a single outburst of SDSS J112003.40+663632.4 to V = 16.9 on 2006 Jan 11 (2). In quiescence, the object is usually below the detection limit of CRTS, although there were two positive detections at V = 21.1 and V = 20.8. We therefore take the quiescence brightness as V ≈ 21, which is also consistent with the SDSS data.

Two of the authors (SB and JS) have monitored the field of SDSS J112003.40+663632.4 on 83 nights between 2009 Jan 1 and 2011 April 10 and during this period have identified 4 separate outbursts as shown in Table 1. In this paper we present photometry from the 2011 Feb outburst from which we conclude that it was a superoutburst. The duration and brightness of the 2010 Apr outburst suggests it too was a superoutburst, although no time resolved photometry was performed to confirm this. Considering that the two outbursts in 2011 were separated by less than 2 months, it is possible that the system undergoes frequent outbursts, but that others have been missed due to incomplete observational coverage.

## Photometry and analysis

Approximately 63 hours of unfiltered photometry was conducted during the 2011 outburst of SDSS J112003.40+663632.4 using the instrumentation shown in Table 2 and according to the observation log in Table 3. Images were dark-subtracted and flat-fielded prior to being measured using differential aperture photometry relative to either GSC 4152-303 (V=13.0) or GSC 4152-583 (V=14.4). Given that the observers used different comparison stars and instrumentation, including CCD cameras with different spectral responses, small systematic differences are likely to exist between observers. However, given that the main aim of our research was to look for time dependent phenomena, we do not consider this to be a significant disadvantage. Nevertheless, where overlapping datasets were obtained during the outburst, we aligned measurements by different observers by experiment. Adjustments of up to 0.09 magnitudes were made. Heliocentric corrections were applied to all data.

## Detection and course of the 2011 Feb outburst

The outburst was detected by SB on 2011 Feb 11 (HJD 2455603.788) at mag 15.6 and the light curve of the outburst is shown in Figure 1. There is a gap of 3 days in the observational record following detection, but the subsequent observations between HJD 2455607 and 2455615 correspond to the plateau phase during which the star gradually faded at a mean rate of 0.12 mag/d. A more rapid fade was observed on HJD 2455615, at a rate of 0.43 mag/d and this may well have been the beginning of the final rapid decline towards quiescence which is characteristic of dwarf novae outbursts. However, the approach to quiescence itself was not observed; observations on HJD 2455619 and 2455620, 16 and 17 days after the outburst was detected, showed that the star was below the detection limit (C > 18.7). Thus the outburst lasted at least 12 days and the amplitude was 5.4 mag above mean quiescence.

## Measurement of the superhump period

We plot expanded views of the longer photometry runs in Figure 2. Superhumps were clearly visible between HJD 2455607 and 2455610 (Figure 2a and b) indicating that this was a superoutburst and that SDSS J112003.40+663632.4 is therefore a member of the SU UMa family of dwarf novae. The amplitude of the superhumps was between 0.17 and 0.22 mag (Table 4, Figure 3a) and did not appear to vary in a systematic way. No superhumps were observed on the night of detection (data not shown), suggesting that we caught the star near the beginning of the superoutburst. Moreover, on the final two nights of photometry, HJD 2455613 and 2455615 (Figure 2c and d), no superhumps were visible, although the data were rather noisy which might well have masked any small superhumps if they had been present.

To study the superhump behaviour, we first extracted the times of each sufficiently well-defined superhump maximum by using the Kwee and van Woerden method (3) in the *Minima v2.3* software (4). Times of 21 superhump maxima were found and are listed in Table 4; in some cases, the same superhump was recorded by two observers, hence two measurements are given. An unweighted linear least squares analysis of the times of maximum allowed us to obtain the following superhump maximum ephemeris:

$$HJD_{max} = + 2455607.40499(89) + 0.07057(19) \times E \qquad \text{Equation 1}$$

Thus the mean superhump period in this interval was $P_{sh}$ = 0.07057(19) d. The observed minus calculated (O–C) residuals for all the superhump maxima relative to the ephemeris are shown in Figure 3b. This suggests that $P_{sh}$ was constant during this stage of the outburst. However, given the rather short time over which the superhumps were observed (3 days) we cannot draw general conclusions about the evolution of $P_{sh}$ during the outburst.

We also carried out a Lomb-Scargle period analysis of the photometry from HJD 2455607 and 2455610, having first subtracted the mean magnitude of the data, using the *Peranso* software version 2.50 (5). The resulting power spectrum (Figure 4) has its highest peak at a frequency of 14.1824(896) cycles/d, which we interpret as the superhump cycle, plus its 1 cycle/d aliases. This correspond to $P_{sh}$ = 0.07051(45) d, which is consistent with the value we obtained from the linear analysis of superhump times. The error estimates are derived using the Schwarzenberg-Czerny method (6). The power spectrum also shows a second group of signals at around 28 cycles/d, corresponding to the second harmonic of the main group. Folding the data on $P_{sh}$ gives the phase diagram show in Figure 5, where 2 cycles are shown for clarity. This shows the typical profile of superhumps, where the rise to maximum is faster than the decline.

Lomb-Scargle period analysis of the photometry from HJD 2455613 and 2455615 did not yield a coherent signal.

**Estimation of $P_{orb}$**

We attempted to identify a signal corresponding to $P_{orb}$ by pre-whitening the Lomb-Scargle power spectrum in Figure 4 with $P_{sh}$, but the only signals remaining were very small, corresponding to the residual superhump signal (data not shown).

Gaensicke *et al.* (7) analysed $P_{orb}$ and $P_{sh}$ for a population of SU UMa systems and derived an empirical relationship between the two values. In the absence of a direct measurement of $P_{orb}$, we used this relationship to estimate $P_{orb}$ = 0.0686(10) d.

**Length of supercycle**

We note that the time between the probable 2010 superoutburst and the one in 2011 is about 295 days. Since the superoutbursts of SU UMa systems are quasi-period, the supercycle is probably around $T_s$ = 295/n days. According to Nogami *et al.* (8), systems with a superoutburst amplitude of 5 to 6 magnitudes tend to have $T_s$ ~200 to 300 days, so n is most likely 1 or 2, giving $T_s$ ~ 150 or ~300 days.

## Conclusions

Unfiltered photometry of SDSS J112003.40+663632.4 during the 2011 Feb outburst revealed the presence of superhumps with peak-to-peak amplitude of up to 0.22 magnitudes, showing this to be an SU UMa type dwarf nova. The outburst amplitude was 5.4 magnitudes above a mean quiescence level of magnitude V ~ 21. The outburst lasted at least 12 days. From a linear analysis of the times of superhump maximum during 3 days of the plateau phase, we measured the superhump period as $P_{sh}$ = 0.07057(19) d. We estimated the orbital period as $P_{orb}$ = 0.0686(10) d. Three further outbursts of the system were recorded between 2009 Jan 1 and 2011 April 10, at least one of which appears to have been another superoutburst. Two outbursts were separated by less than 2 months, which suggests that the system may undergo frequent outbursts

We urge further monitoring of SDSS J112003.40+663632.4 with the aim of identifying the frequency of normal outbursts and superoutbursts.

## Acknowledgements

The authors gratefully acknowledge the use of data from the Catalina Real-Time Transient Survey. We also used SIMBAD, operated through the Centre de Données Astronomiques (Strasbourg, France) and the NASA/Smithsonian Astrophysics Data System. We thank the referees, Dr. Chris Lloyd and Dr. Robert Smith, for their helpful comments that have improved the paper.

## Addresses

JS: "Pemberton", School Lane, Bunbury, Tarporley, Cheshire, CW6 9NR, UK [bunburyobservatory@hotmail.com]

SB: 5 Melba Drive, Hudson, NH 03051, USA [sbrady10@verizon.net]
TK: CBA New Mexico, PO Box 1351 Cloudcroft, New Mexico 88317, USA [tom_krajci@tularosa.net]

EdM: Departamento de Fisica Aplicada, Facultad de Ciencias Experimentales, Universidad de Huelva, 21071 Huelva, Spain; Center for Backyard Astrophysics, Observatorio del CIECEM, Parque Dunar, Matalascañas, 21760 Almonte, Huelva, Spain [demiguel@uhu.es]

MP: 3206 Overland Ave, Baltimore, MD 21214 USA [mike@orionsound.com]

RS: 2336 Trailcrest Dr., Bozeman, MT 59718, USA [richard@theglobal.net]

WS: 6025 Calle Paraiso, Las Cruces, NM 88012, USA [starman@tbelc.org]

| Detection date (UT) | Maximum brightness (mag) | Duration (d) | Comment |
|---|---|---|---|
| 2009 Nov 29 | 16.4 | >1 | Only detected on a single night |
| 2010 Apr 22 | 16.3 | >10 | |
| 2011 Feb 11 | 15.6 | >12 | Current paper |
| 2011 Apr 2 | 16.5 | >2 | |

**Table 1: Observed outbursts of SDSS J112003.40+663632.4 between 2009 Jan 1 and 2011 April 10**

| Observer | Telescope | CCD |
|---|---|---|
| Brady | 0.4 m reflector | SBIG ST-8XME |
| Krajci | 0.3 m SCT | SBIG ST9-XME |
| de Miguel | 0.28 m SCT | QSI-516ws |
| Potter | 0.35 m SCT | SBIG ST-10XME |
| Sabo | 0.43 m reflector | SBIG STL-1001 |
| Shears | 0.28 m SCT | Starlight Xpress SXVF-H9 |
| Stein | 0.35 m SCT | SBIG ST-10XME |

**Table 2: Instrumentation**

| Start date (UT) | Start time (HJD) | End time (HJD) | Duration (h) | Observer |
|---|---|---|---|---|
| 2011 Feb 11 | 2455603.788 | 2455603.938 | 3.6 | Brady |
| 2011 Feb 14 | 2455607.346 | 2455607.509 | 3.9 | Shears |
| 2011 Feb 15 | 2455607.664 | 2455607.851 | 4.5 | Sabo |
| 2011 Feb 15 | 2455607.864 | 2455608.023 | 3.8 | Stein |
| 2011 Feb 15 | 2455607.875 | 2455608.042 | 4.9 | Krajci |
| 2011 Feb 15 | 2455608.321 | 2455608.649 | 7.9 | de Miguel |
| 2011 Feb 16 | 2455608.508 | 2455608.712 | 4.9 | Potter |
| 2011 Feb 16 | 2455608.848 | 2455609.025 | 4.2 | Stein |
| 2011 Feb 16 | 2455608.875 | 2455608.905 | 0.7 | Krajci |
| 2011 Feb 17 | 2455609.875 | 2455610.041 | 4.0 | Krajci |
| 2011 Feb 18 | 2455610.323 | 2455610.438 | 2.8 | de Miguel |
| 2011 Feb 20 | 2455613.324 | 2455613.613 | 6.9 | de Miguel |
| 2011 Feb 22 | 2455615.493 | 2455615.839 | 8.3 | Potter |
| 2011 Feb 22 | 2455615.498 | 2455615.615 | 2.8 | Brady |
| 2011 Feb 26 | 2455619.440 | 2455619.447 | 0.17* | Shears |
| 2011 Feb 27 | 2455620.449 | 2455620.456 | 0.17* | Shears |

**Table 3: Observation log**

| Superhump cycle number | Superhump maximum (HJD) | Error (d) | O-C (d) | Superhump amplitude (mag) |
|---:|---|---|---|---|
| 0 | 2455607.4047 | 0.0012 | -0.0003 | 0.19 |
| 1 | 2455607.4777 | 0.0012 | 0.0021 | 0.18 |
| 4 | 2455607.6879 | 0.0024 | 0.0006 | 0.22 |
| 5 | 2455607.7575 | 0.0021 | -0.0003 | 0.22 |
| 6 | 2455607.8293 | 0.0015 | 0.0009 | 0.21 |
| 7 | 2455607.9001 | 0.0012 | 0.0011 | 0.22 |
| 7 | 2455607.9010 | 0.0024 | 0.0020 | 0.21 |
| 8 | 2455607.9698 | 0.0009 | 0.0003 | 0.21 |
| 8 | 2455607.9684 | 0.0015 | -0.0011 | 0.22 |
| 14 | 2455608.3908 | 0.0045 | -0.0022 | 0.19 |
| 15 | 2455608.4660 | 0.0048 | 0.0025 | 0.18 |
| 16 | 2455608.5333 | 0.0042 | -0.0008 | 0.20 |
| 16 | 2455608.5319 | 0.0015 | -0.0022 | 0.19 |
| 17 | 2455608.6035 | 0.0042 | -0.0012 | 0.20 |
| 17 | 2455608.6030 | 0.0015 | -0.0017 | 0.17 |
| 18 | 2455608.6731 | 0.0024 | -0.0022 | 0.19 |
| 21 | 2455608.8858 | 0.0018 | -0.0012 | 0.19 |
| 22 | 2455608.9578 | 0.0024 | 0.0003 | 0.20 |
| 36 | 2455609.9461 | 0.0018 | 0.0006 | 0.18 |
| 37 | 2455610.0177 | 0.0018 | 0.0016 | 0.18 |
| 42 | 2455610.3698 | 0.0012 | 0.0009 | 0.22 |

**Table 4: Times and amplitudes of superhumps**

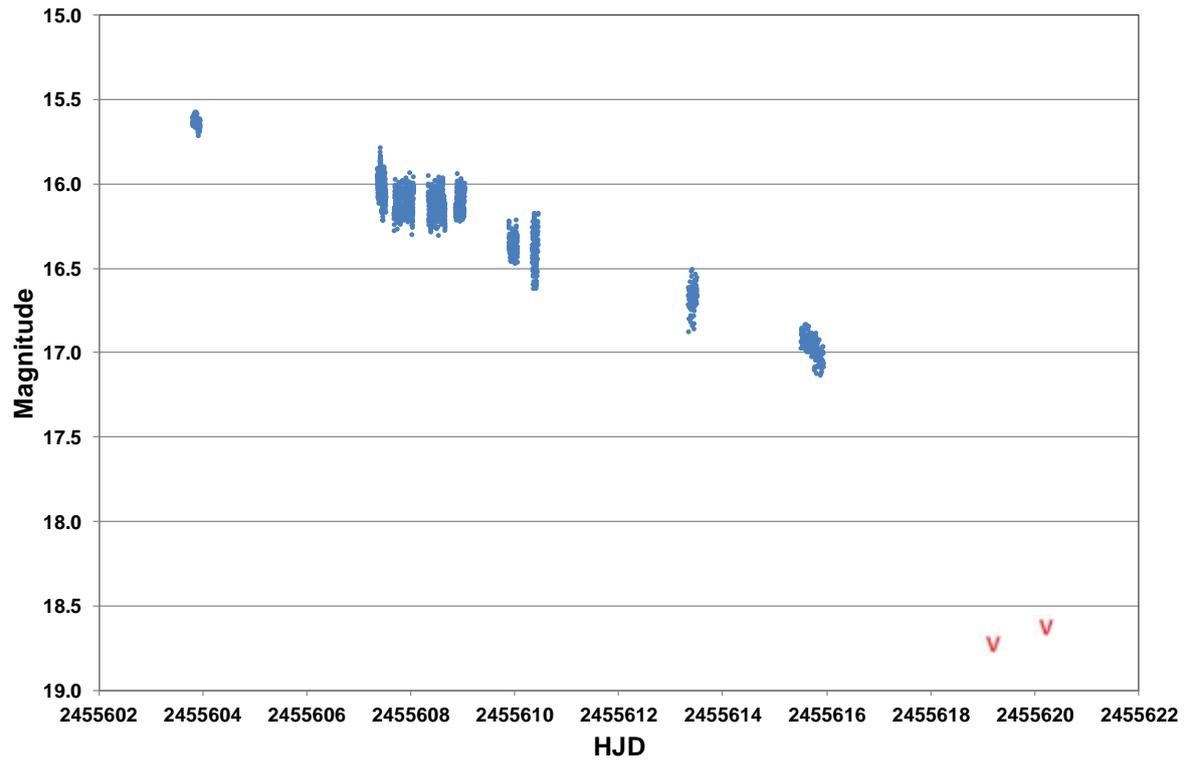

**Figure 1: Outburst light curve**

**V**= upper limit ("fainter than") observations

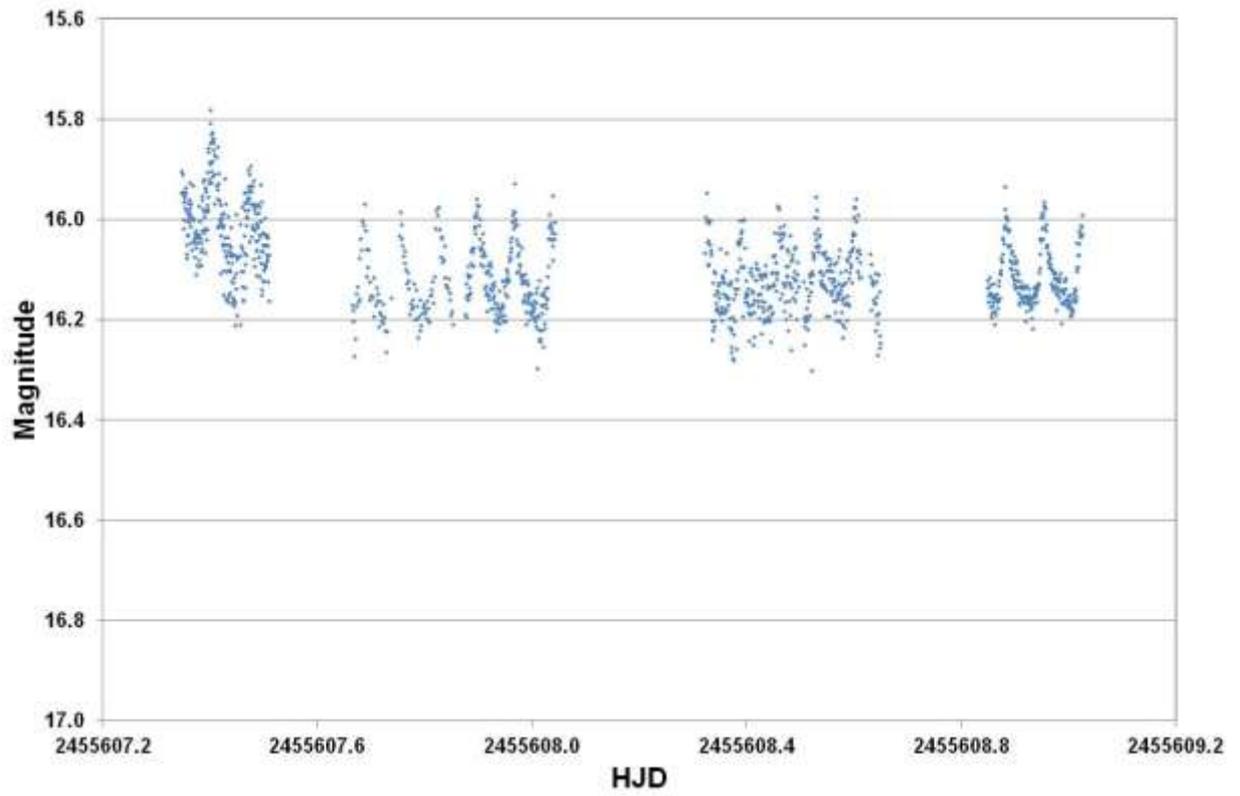

**(a)**

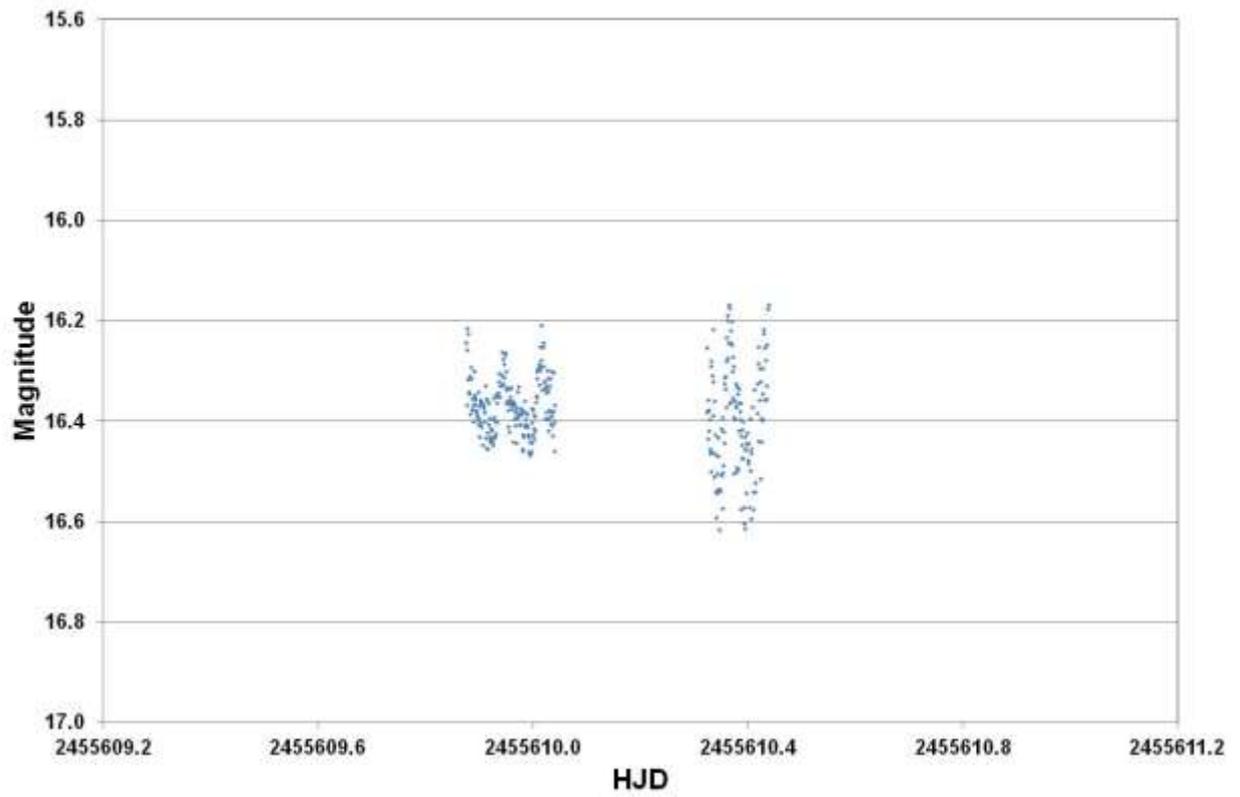

**(b)**

*Accepted for publication in the Journal of the British Astronomical Association*

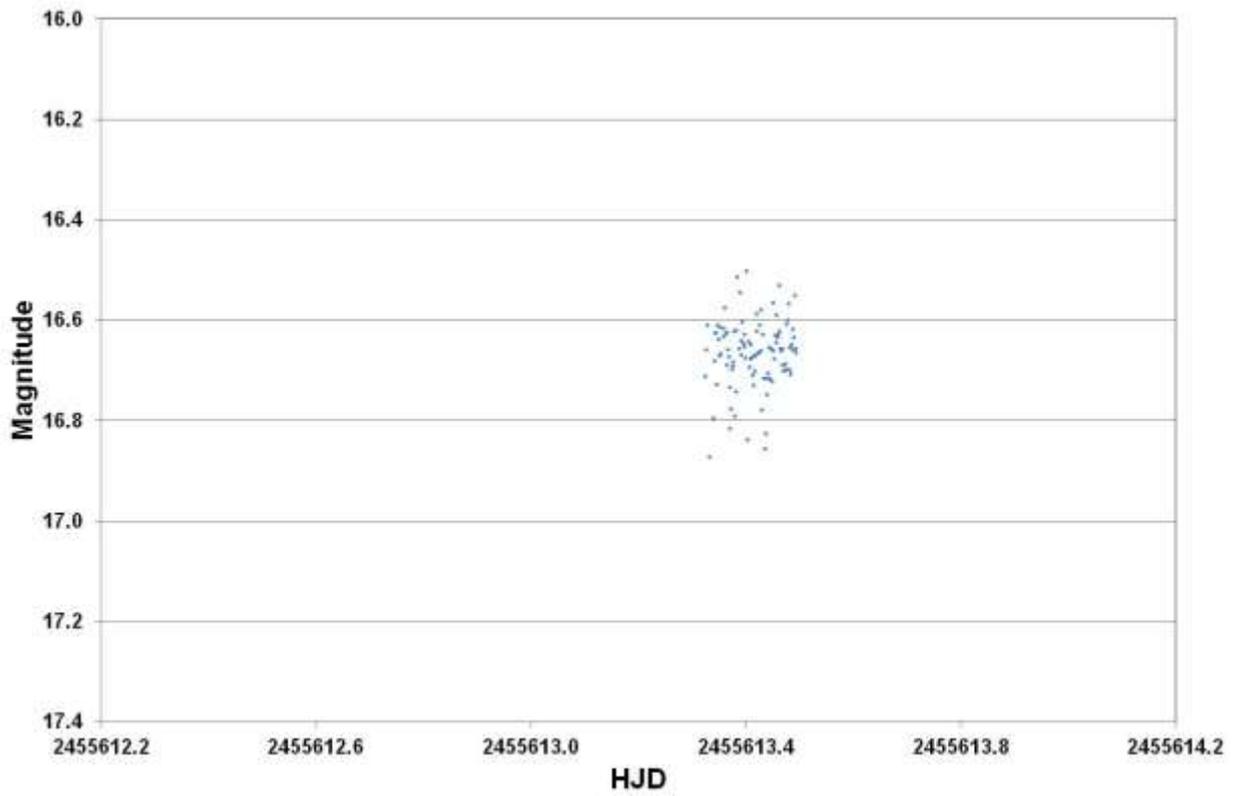

(c)

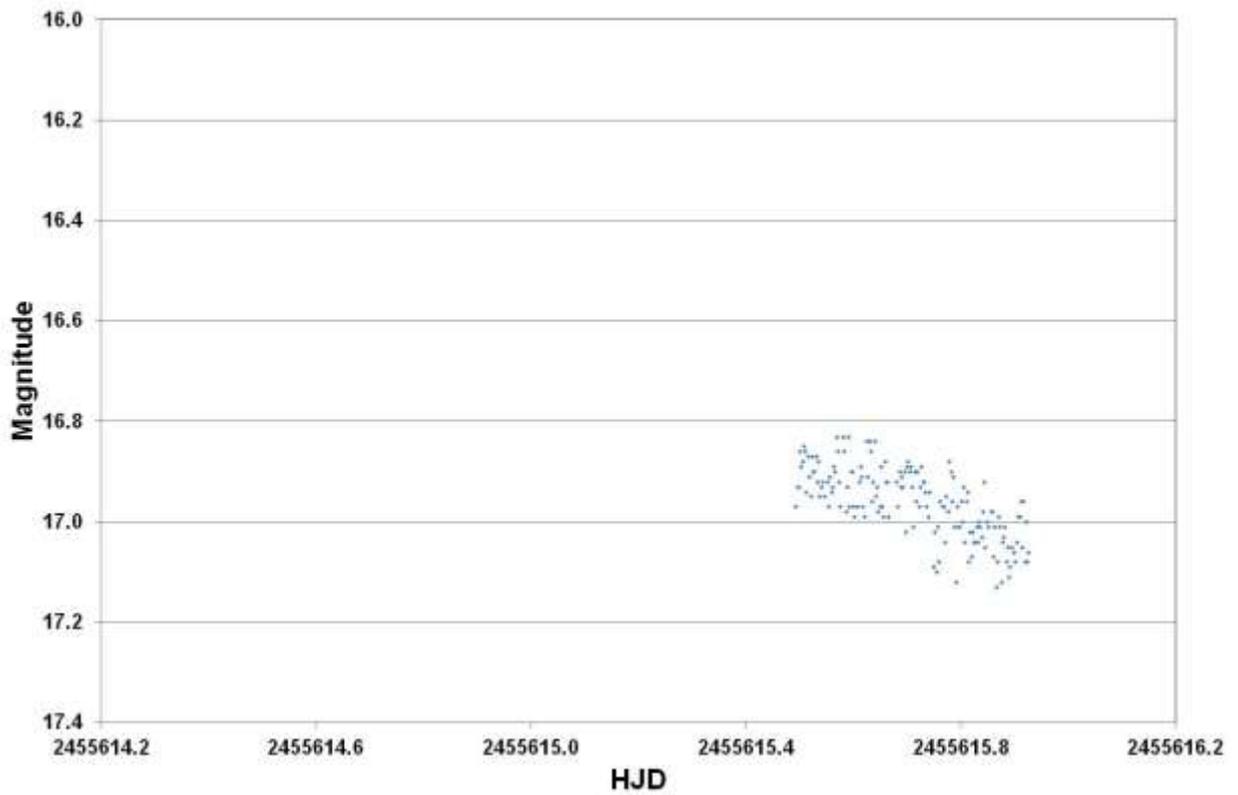

(d)

**Figure 2: Time resolved photometry**

*Accepted for publication in the Journal of the British Astronomical Association*

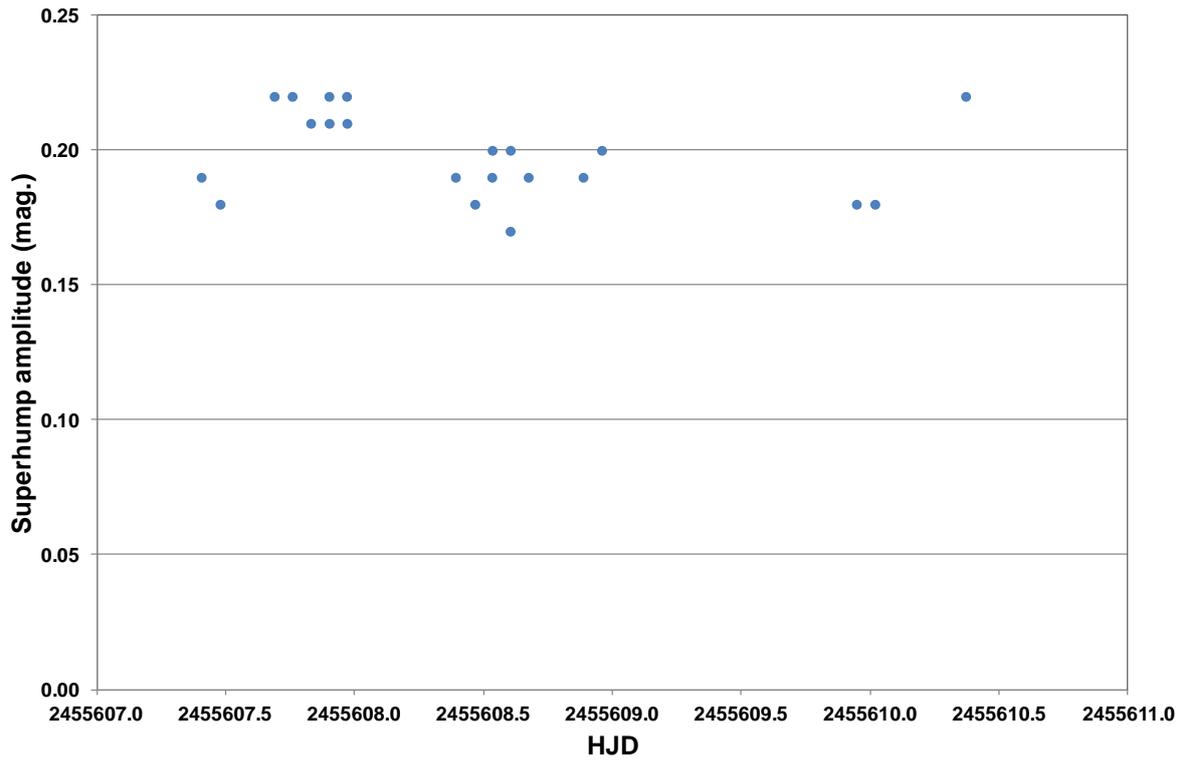

(a)

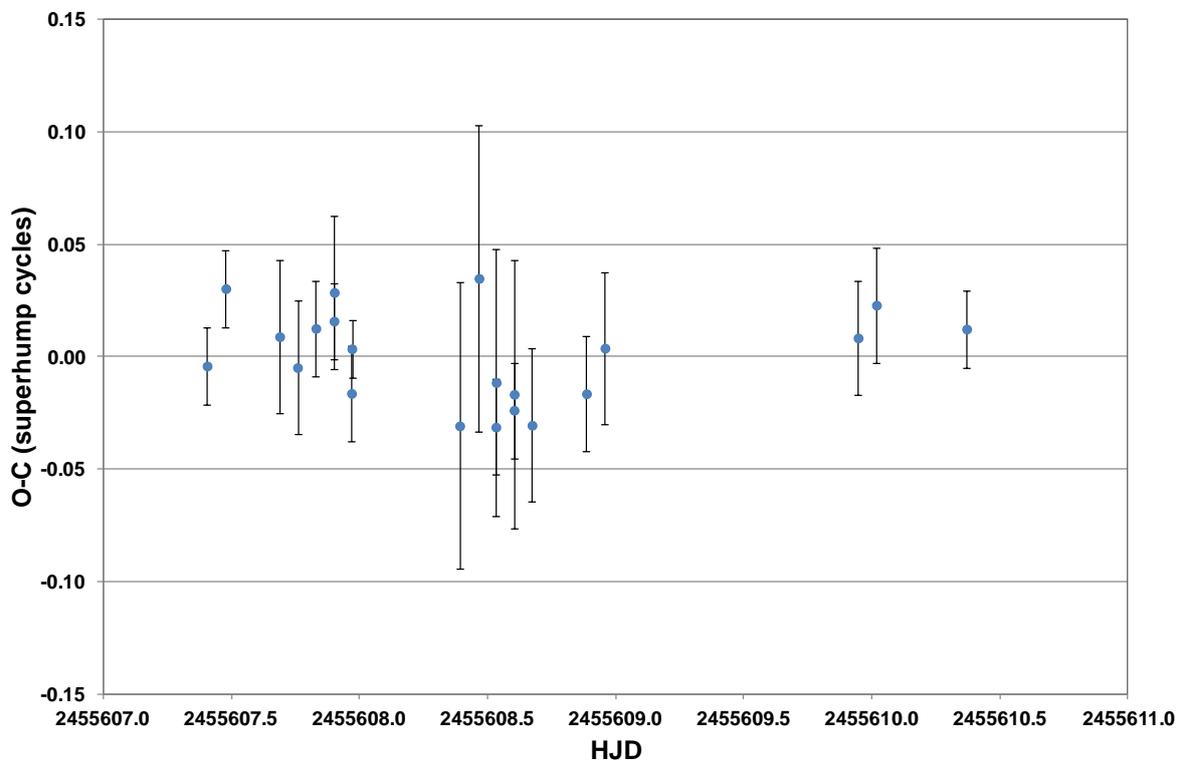

(b)

**Figure 3: Amplitude (a) and O-C diagram (b) of the superhumps**

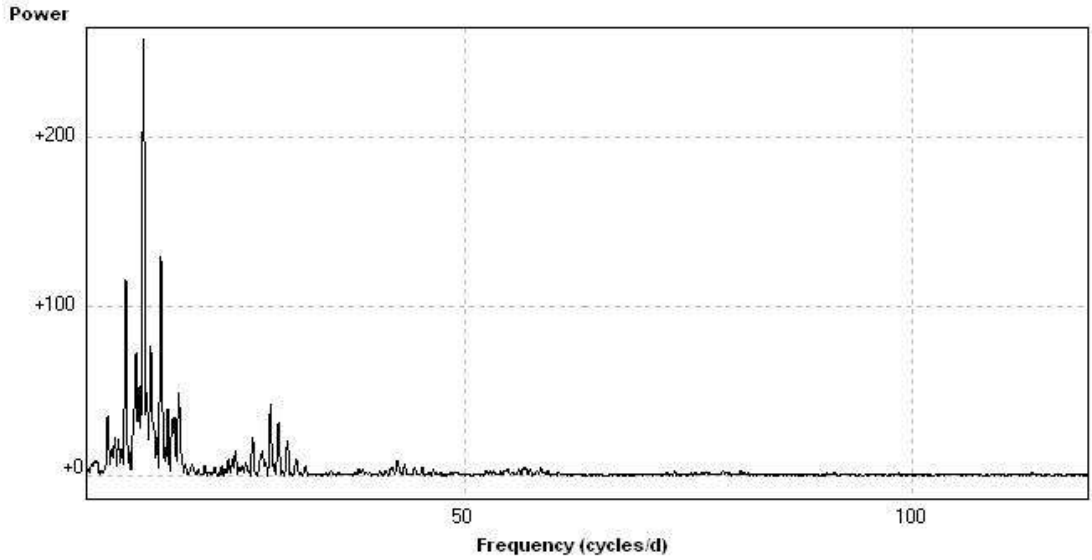

**Figure 4: Lomb-Scargle power spectrum**

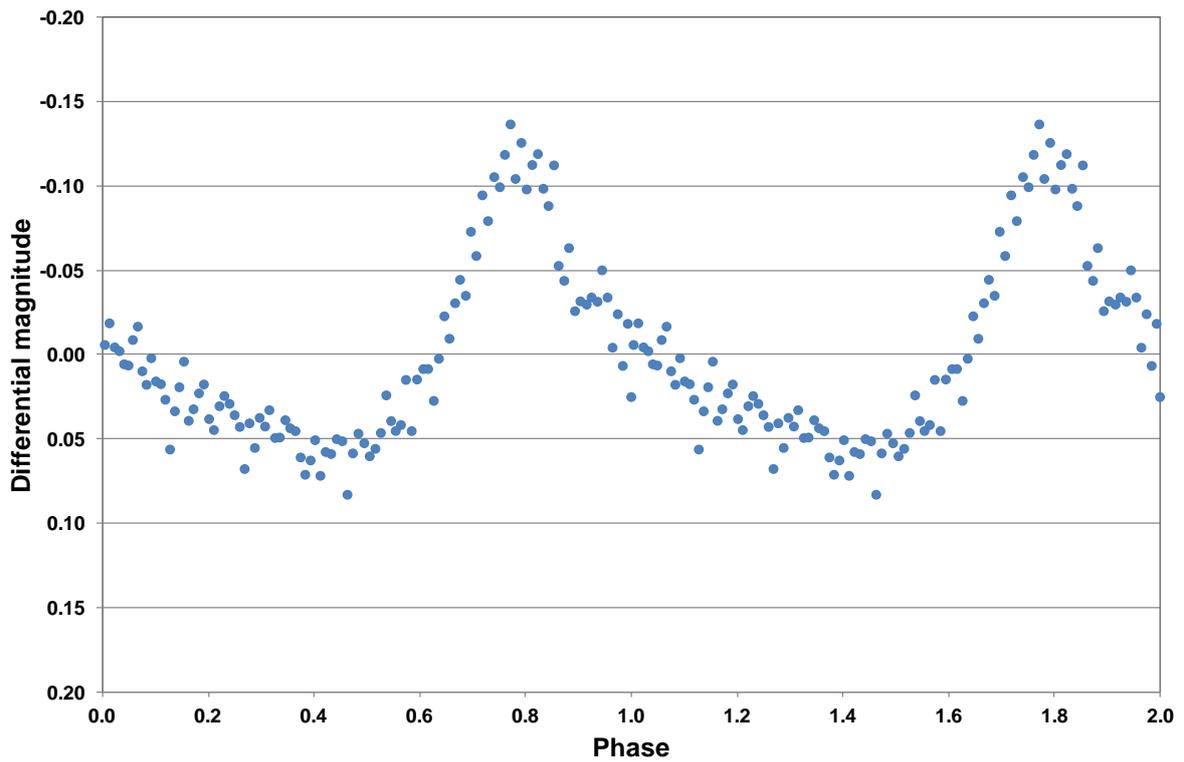

**Figure 5: Phase diagram of the superhumps folded on $P_{sh}$ = 0.07051 d**

Each data point is the mean of 10 individual measurements